\documentstyle[12pt]{article}
\newcommand{\rf}[1]{(\ref{#1})}
\newcommand{\bea}{\begin{eqnarray}}
\newcommand{\eea}{\end{eqnarray}}
\newcommand{\g}{\gamma}

\renewcommand{\b}{\beta}
\renewcommand{\a}{\alpha}

\newcommand{\sg}{\sigma}
\newcommand{\k}{\kappa}
\newcommand{\vph}{\varphi}
\newcommand{\oh}{\frac{1}{2}}

\newcommand{\ra}{\right\rangle}
\newcommand{\la}{\left\langle}

\newcommand{\cT}{{\cal T}}

\def\void{}
\def\labelmark{}

\newenvironment{formula}[1]{\def\labelname{#1}
\ifx\void\labelname\def\junk{\begin{displaymath}}
\else\def\junk{\begin{equation}\label{\labelname}}\fi\junk}%
{\ifx\void\labelname\def\junk{\end{displaymath}}
\else\def\junk{\end{equation}}\fi\junk\labelmark\def\labelname{}}

{\ifx\void\labelname\def\junk{\end{array}\end{displaymath}}
\else\def\junk{\end{array}\right.\end{equation}}
\fi\junk\labelmark\def\labelname{}\def\junk{}
}

\newcommand{\beq}{\begin{formula}}
\newcommand{\eeq}{\end{formula}}
\newcommand{\beqv}{\begin{formula}{}}

\begin{document}
\topmargin 0pt
\oddsidemargin 5mm
\headheight 0pt
\headsep 0pt
\topskip 9mm

\hfill    NBI-HE-93-3

\hfill January 1993

\begin{center}
\vspace{24pt}
{\large \bf 4d quantum gravity coupled to matter}
\vspace{24pt}

{\sl J. Ambj\o rn }

\vspace{6pt}

The Niels Bohr Institute\\
Blegdamsvej 17, DK-2100 Copenhagen \O , Denmark\\

\vspace{12pt}

\sl Z. Burda\footnotemark[1] and J. Jurkiewicz\footnote[1]
{Supported by the KBN grants no. 2 0053 91 01 and 2P302-16904}

\vspace{6pt}

Inst. of Phys., Jagellonian University., \\
ul. Reymonta 4, PL-30 059, Krak\'{o}w~16, Poland

\vspace{12pt}

{\sl C.F. Kristjansen}

\vspace{6pt}

 The Niels Bohr Institute\\
Blegdamsvej 17, DK-2100 Copenhagen \O , Denmark\\

\end{center}

\vspace{24pt}

\begin{center}
{\bf Abstract}
\end{center}

\vspace{12pt}

\noindent
We investigate the phase structure of four-dimensional quantum gravity
coupled to Ising spins or Gaussian scalar fields
by means of numerical simulations.
The quantum gravity part is modelled by the summation over random
simplicial manifolds, and the matter fields are located in the center
of the 4-simplices, which constitute the building blocks of the
manifolds. We find that the coupling between spin and geometry
is weak away from the critical point of the Ising model. At the
critical point there is clear coupling, which qualitatively agrees with
that of gaussian fields coupled to gravity. In the case of pure gravity
a transition between a phase with highly connected geometry
and a phase with very ``dilute'' geometry has been observed earlier.
The nature of this transition seems unaltered when matter fields are
included. It was the hope that
continuum physics could be extracted at the transition between
the two types of
geometries. The coupling to matter fields, at least in the form
discussed in this paper, seems not to improve the
scaling of the curvature at the transition point.



\newpage
\addtolength{\baselineskip}{0.20\baselineskip}

\section{Introduction}

Last year a new regularized model of quantum gravity in 4D was introduced
\cite{am,aj}. The path integral is approximated by a summation
over randomly triangulated piecewise
linear manifolds\footnote{An older, related approach makes
use of a fixed triangulation, but allows the variation of the
length of the links. Contrary, in the present approach one keeps the length
of the links fixed, but varies the connectivity. We refer to
\cite{hamber} for a recent lucid review of the first approach, which
we here will call ``Regge gravity'', while we will use the term ``simplicial
gravity'' for the present approach.}. This method is a  generalization of
the one from two dimensions, which was very successful
\cite{david1,adf,kkm,david2}.
In 4D simplicial quantum gravity two
different phases have been observed, one with a highly
connected geometry and a large Hausdorff dimension and one with
a low Hausdorff dimension. Based on numerical simulations it was suggested
in \cite{aj} that the transition between the
two types of geometries was of second order
and that an interesting continuum limit might be extracted at the transition
point. This observation has been further corroborated in a sequence of
papers \cite{varsted,am1,ajk,bruegmann,bm}.

One obstacle to the above mentioned suggestion  is that
the average curvature does not scale to zero at the transition point.
The average curvature does decrease (albeit slowly) with the volume of
the simulated universes and it cannot be completely
ruled out that it actually scales
to zero in the infinite volume limit. However, at the moment we
consider it as unlikely. This prompts at least a reinterpretation of
the meaning of the scaling limit since naive scaling like
\beq{*1}
\la R_{lattice}\ra = \la R_{cont}\ra  a^2
\eeq
(where $a$ is the lattice spacing)
cannot be maintained. Maybe the average curvature should be absorbed in a
redefinition of the cosmological constant, while the relevant physical
curvature arises only through fluctuations around the ``fictitious''
average curvature. While such an unconventional limit might exist,
it seems not to be very natural to us.  An attempt
to improve the situation by adding terms like $R^2$ to the action
was not  very successful~\cite{ajk}.
At this point we should mention a recent suggestion
\cite{amm} of a different identification of the lattice results with
continuum theory in which one considers the limit of the bare gravitational
coupling constant
going to infinity. This limit might in continuum language correspond
to an infrared fixed point dominated by the quantum fluctuations
of the conformal factor. The scaling relations derived in \cite{amm}
agree at the qualitative level quite well with the numerical results,
but they move the interesting region of continuum physics
away from the transition in geometry and to a region in coupling
constant space where \rf{*1} can be satisfied. We consider this suggestion
as most interesting. In this article we
explore another way to cure the problem with the  scaling of the
average curvature, namely coupling of matter fields to gravity. It
is of course also of interest in itself to study matter fields coupled
to dynamical random geometries. In the best of all worlds one could even
hope that the quest for correct scaling of gravity observables like
the average curvature would uniquely determine the matter content of the
theory\footnote{But we will of course not seriously pretend, that the present
stage of numerical simulations of quantum gravity is such, that one
could really determine the matter content.}.

The coupling of matter to two-dimensional gravity has revealed a
rich and beautiful structure as long as the central charge of the field
theory is less than or equal to one. This is summarized in the KPZ
formulas \cite{kpz}, but was first discovered in the
simplicial gravity approach.
As an example, when the
Ising model is coupled to 2d simplicial gravity its phase transition
changes from being second order
to third order \cite{kazakov,bk}.
In addition the back-reaction of matter changes
the critical exponent $\g$ of gravity {\it at} the critical point
of the Ising model. Away from the critical point this exponent
is unchanged.

Unfortunately the analytical methods of 2d have not yet been extended to
higher dimensions. The coupling of the Ising model to 3d gravity
was investigated by numerical simulations in \cite{baillie,cr,abjk}.
The phase diagram was determined in \cite{abjk} and the conclusion was
that, although there was a clear coupling between gravity and the spins
at the critical point of the spin system, this influence was
not sufficiently strong to change the  {\it first order transition}
observed in three dimensions \cite{abkv,av} between the two phases of
the geometrical system into a more interesting (from the point of
view of continuum physics) second order transition.
In this respect the situation is better in 4d  where the
transition between the two phases of the geometrical system
may already be of second order, as mentioned above.

The rest of this paper is organized as follows:
In section \ref{model} we define the model. In section \ref{method}
we discuss briefly the numerical method, while section \ref{results}
contains our numerical results. Finally in section \ref{discuss}
we discuss the results obtained.

\section{The model \label{model}}

Simplicial quantum gravity in 4d is described by the following partition
function (see e.g. \cite{aj,ajk}):
\beq{*2}
Z(\k_2,\k_4)=\sum_{T\in \cT} e^{-\k_4 N_4 +\k_2 N_2}
\eeq
where the sum is over triangulations $T$ in a suitable class of
triangulations $\cT$.
The quantity
$N_4$ denotes the number of 4-simplexes in the triangulation
and $N_2$ the number of triangles. The coupling constant
$\k_2$ is inversely proportional to the bare gravitational coupling
constant, while $\k_4$ is related to the bare cosmological constant.
The most important restriction
to be imposed on $\cT$ is that of a fixed topology. If we allow an
unrestricted summation over all topologies in \rf{*2} the partition function
is divergent \cite{aj}. In the following we will always restrict ourself
to consider manifolds with the topology of $S^4$.

$Z(\k_2,\k_4)$ is the grand canonical partition function.
It is defined in a region
$\k_4 \geq \k_4^c(\k_2)$ in the $(\k_2,\k_4)$ coupling constant plane.
The only way in which we can hope to obtain a continuum limit is by
letting $\k_4$ approach $\k_4^c(\k_2)$
from above. This tentative continuum limit depends only on
one coupling constant $\k_2$ and the
transition between the two phases of 4d gravity mentioned
above takes place at a critical value of $\k_2$, $\k_2^c$.
It is often convenient to think about the canonical partition function
where $N_4$ is kept fixed. Then $\k_2$ is the only coupling constant
and the aspects of gravity which do not involve the fluctuation of the
total volume of the universe can be addressed in the limit of large $N_4$.
For the geometrical system an observable which has our interest
is the average curvature per volume, $\la R \ra$.
The average curvature
can for a simplicial manifold be defined by Regge calculus and in the
case of equilateral simplexes one simply has
\beq{*ac}
\la R \ra \propto  (c_4 N_2/N_4 -10)
\eeq
where the constant $c_4$ is the number of 4-simplexes to which each
triangle should belong if the manifold were flat.
Furthermore one can by an appropriate interpretation of the Regge approach
introduce the average of the squared curvature per volume by
\beq{*r2}
\la R^2 \ra \propto
\frac
{\sum_{n_2} o(n_2)\left[(c_4-o(n_2))/o(n_2)\right] ^2}{10N_4}
\eeq
where the sum is over triangles $n_2$ and $o(n_2)$ is the order of a
given triangle i.\ e.\ the number of 4-simplexes to which this triangle
belongs.
The correlator $\la R^2 \ra - \la R \ra ^2$ will prove useful as
an indicator of a change in geometry.

One can now couple matter fields to simplicial quantum gravity. In the
case of Ising spins the partition function will look like:
\beq{*3}
Z(\b,\k_2,\k_4)=\sum_{N_4} e^{-\k_4 N_4}
\sum_{T \in \cT (N_4)}\sum_{\{\sg\}}e^{\k_2 N_2}
e^{\b \sum_{\la i,j\ra} (\delta_{\sg_i \sg_i}-1)}.
\eeq
In this formula $\cT(N_4)$ signifies the subclass of $\cT$ with volume $N_4$,
$\sum_{\{\sg\}}$ the summation over all spin configurations, while
$\sum_{\la i,j\ra}$ stands for the summation over all neighbouring pairs of
4-simplexes. As a function of $\b$ there might or might not be a phase
transition for the spin system, depending on the value of $\k_2$ (assuming that
$\k_4= \k_4^c(\k_2,\b)$, where $\k_4^c$ now depends on both $\k_2$ and $\b$).

The coupling of scalar fields to simplicial quantum gravity is also
straightforward. Here we will ignore self-interaction of the
scalar fields and direct coupling between the scalar fields and the
curvature, and simply consider the following partition function
\beq{*4}
Z(\k_2,\k_4) =  \sum_{N_4} \sum_{T \in \cT (N_4)}e^{\k_2 N_2-\k_4N_4}
\int \prod_{i,\a}\frac{d\phi_i^\a}{\sqrt{2\pi}} \;
\prod_{\a=1}^{n_g}\delta (\sum_{i} \phi_i^\a)
\;e^{ -\oh\sum_{\la i,j\ra,\a} (\phi_i^\a-\phi_j^\a)^2}.
\eeq
Here $i$ labels the 4-simplexes, $\a$ different components of
the field $\phi$
and $n_g$ is the total number of independent Gaussian fields. There is no
need for a coupling constant in front of the Gaussian action since it
can always be absorbed in $\k_4$ by a rescaling of the $\phi$'s.
Of course the gaussian action can in principle be integrated out explicitly,
leaving us with an additional weight
\beq{*5}
\left({\rm Det}\;C_T\right)^{-n_g/2}
\eeq
for each triangulation $T$, where $C_T$ is just the incidence matrix for the
$\vph^5$-graph which is dual to the triangulation $T$. In the case of
gaussian fields coupled to 2d gravity this fact was used to determine
qualitatively the phase diagram of non-critical strings as
a function of the number of Gaussian systems,
$n_g$ \cite{adfo,adf1,bkkm}.
In principle one
could try to do the same here. However, the class of allowed $\vph^5$
graphs is not so easy to determine  as in the case of 2d gravity. In the
following we will rely on numerical simulations.

\section{Numerical methods\label{method}}

One annoying aspect
of the above formalism is that we are forced to perform a grand canonical
simulation where $N_4$ is not fixed. The reason is that we
have no ergodic updating algorithm\footnote{In 2d
gravity we know how to perform
a canonical updating, but even there the grand canonical updating is
occasionally convenient to use \cite{jkp,adfo,adf1,afkp}.} which preserves
the volume $N_4$. It is however possible to perform a grand canonical
updating without violating ergodicity
and  still stay in the neighbourhood of a prescribed value
of $N_4$, which we will denote $N_4(fix)$. The procedure involves
finetuning of $\k_4$ to its critical value, $\k_4^c(\k_2,\beta)$.
We refer to \cite{ajk} for details.

In addition to the updating of the geometry, we also have to update the
Ising spin system and the Gaussian systems.
Let us first discuss the Ising spin system.
In order to avoid critical slowing down close to the
phase transition between the magnetized and the non-magnetized phase
the spin  updating is performed by  the single
cluster variant of the Swendsen-Wang algorithm developed by Wolff \cite{wolf}.
The cluster updating algorithms have been successfully applied to
the Ising model coupled to 2d gravity \cite{jj,bj,bj1,rk}
and to the Ising model coupled to 3d gravity \cite{abjk}.
We update the spins once for every sweep, i.e. after $N_4(fix)$ {\it accepted}
updatings of the geometry.

In the simulations we have scanned the $(\k_2,\b)$ coupling constant
plane by first fixing $\k_2$ and then varying $\b$ in the search for
a critical value $\b_c(\k_2)$ where the spin system undergoes
a transition\footnote{In order that the reader could appreciate the
amount of work going into this please note that we  have to fine-tune
$\k_4$ for each value of $\k_2$ and $\b$.}.
For values of $\k_2$ where we are well inside the phase with a highly
connected geometry where
and a large Hausdorff dimension, 5000 sweeps are sufficient
to achieve equilibrium for bulk quantities when the number of simplexes
does not exceed $N_4=9000$. This is in agreement
with the situation in pure gravity \cite{aj,ajk}.
We have occasionally made
longer runs  in connection with the
measurement of Binders cumulant (50.000 sweeps) and
near critical points either in the spin or gravity coupling
constant. It seems as if the situation is in all respects  as in
2d and 3d gravity. In particular the presence of the spins seems not to
slow down the convergence of bulk geometric observables (in 2d it is
known that spins speed it up). In this phase we have neither seen
excessive signs of autocorrelations of spins (the longest of the order
of 500 sweeps at the spin transition).
This is in agreement with intuition since the connectivity of the
system is large and the maximal distance between spins correspondingly
small. The situation is somewhat different when we probe the phase where the
geometry is elongated and
where internal distances can be quite large. Without spin the
convergence in geometry is slow in this phase and it is true also after
coupling to spins.

\vspace{12pt}
\noindent
The Gaussian fields are updated by a heatbath algorithm.
There are two aspects of this updating. One type of updating
is performed with a fixed background geometry and is standard.
The other one is related to the Metropolis updating of the
geometrical structure. Since there are slightly
 unconventional aspects connected with the change of the fields,
 when the geometry is changed \footnote{The same
 aspect is already present in the
grand canonical algorithms used in 2d gravity, see e.g. \cite{jkp}.},
let us make a few comments. We will not go into
details (which are trivial, but clumsy
to write down explicitly), but rather sketch the main point: Consider
a change in geometry where we take a 4-simplex, remove the ``interior'',
insert a vertex in the ``empty'' interior and connect this vertex to
the five vertices of the former 4-simplex. With a
proper identification of sub-simplexes we have by this procedure
removed one 4-simplex and created five new ones. The inverse ``move''
is one where we remove a vertex of order five and
the associated five 4-simplexes and replace them by a single 4-simplex.
We must be careful to treat the Gaussian fields correctly in
such moves.
In the case where we insert a vertex we will have to introduce
five new fields $\vph_i$, $i=1,\ldots 5$. They will interact quadratically with
each other, and each of them will interact with one field
associated with a neighbouring 4-simplex untouched by the move. Let us
denote these five fields $\phi_i$, $i=1\ldots 5$
 In addition we have removed a field
associated with the original 4-simplex. We denote it by $\vph_0$. It
interacted with the five $\phi_i$'s. The correct probability distribution
of the new five $\vph_i$'s is
\beq{*5a}
dP_{new}(\vph_i)= C_{new}(\phi_i)
\prod_{i=1}^5 d\vph_i\; e^{- S_{new}(\vph_i,\phi_i)}
\eeq
where the additional part of the action $S_{new}$ coming from
added fields $\vph_i$, determined from \rf{*4} is
\beq{*5b}
S_{new}(\vph_i)= \oh \sum_{i<j}(\vph_i-\vph_j)^2+\oh\sum_i (\vph_i-\phi_i)^2
\eeq
The factor $C(\phi_i)$ is a normalization factor, which contains the
exponential of a quadratic form in the $\phi_i$'s and its all-over
scale is fixed by the
requirement that $\int dP_{new}(\vph_i) =1$.
In a similar way the field $\vph_0$
which was removed had a Gaussian probability distribution $dP_{old}(\vph_0)$,
just with another action
\beq{*5c}
S_{old}(\vph_0) = \oh \sum_i (\vph_0 -\phi_i)^2
\eeq
and an appropriate normalization factor $C_{old}(\phi_i)$, which again
contains the exponential of a Gaussian form in $\phi_i$'s. Assuming that
the fields $\vph_0,\ldots,\vph_5$ are selected according to $P_{new}$ and
$P_{old}$ it is easy to enlarge the condition for detailed balance for
the change in geometry to include the additional change in field content.

The geometrical moves fall in three classes (see e.g. \cite{aj} for details)
of which we have described one above. A second class is one where
two neighbouring 4-simplexes are removed and replaced by three new ones
having in common a link (a 1-simplex), or the inverse move, where
three 4-simplexes sharing a link are removed and replaced by
two 4-simplexes being neighbours (i.e. sharing a 3-simplex). Finally
the third class of moves is ``self-dual'': three 4-simplexes sharing
a triangle (a 2-simplex) are replaced by three others, sharing
a different triangle. In all cases one can easily write down
$dP_{new}$ and $dP_{old}$ as above and incorporate these probabilities
in the requirement of detailed balance needed for performing
the purely geometrical move.

The total updating is now organized in the following way: A sweep
over the lattice with an updating of geometry and the above described
updating of field content is followed by a number of sweeps with the
geometry fixed and ordinary heatbath updating of the Gaussian fields.
The actual number of such heatbath updatings for each geometrical updating
is chosen so that the fastest convergence to equilibrium is achieved.
For one gaussian field two heatbath updatings for each geometrical
updating is usually sufficient as long as the geometry is highly
connected. In the elongated phase up to 15 gaussian updatings were needed.
The number of necessary updatings per sweep increases with the number of
Gaussian fields. For 4 Gaussian fields 3 updatings per sweep were needed
in the highly connected phase of gravity.

\section{Numerical results \label{results}}

\subsection{Ising spins coupled to gravity}

Pure 4d gravity has two phases and this fact is not changed by the coupling
to a single Ising spin.

In the phase where the geometry is highly connected the spin system
has a phase transition. In fig. 1 we show the absolute value of the
magnetization
\beq{*6}
|\sg| = \frac{1}{N_4} \left|\sum_{i=1}^{N_4} \sg_i \right|
\eeq
as a function of $\b$ for a value of $\k_2$ for which the geometrical
system is highly connected. In fig.2 we show Binders cumulant
defined by
\beq{*6a}
B(\b) = 1- \frac{1}{3} \frac{\la \sg^4 \ra}{\la \sg^2 \ra^2}
\eeq
and it is seen
that the data are consistent with a transition which is second order or
higher. We feel there is no reason to believe that the  transition
should be of higher than second order, since in this phase of
the geometrical system the effective Hausdorff
dimension is quite high which should favour mean-field results.
In the phase with elongated geometry the situation is quite different.
The magnetization curve well inside this phase is shown in fig.3.
There is only a gradual cross over to $|\sg| \approx 1$ for large $\b$, and
the cross over weakens (slightly)
with increasing volume. This is in agreement with
the measurements of the Hausdorff dimension, $d_H$, in this phase which seems
to indicate that $d_H < 2$.

The phase diagram in the $(\k_2,\b)$ plane is as it appears
for a system consisting of 9K simplexes is shown in fig.4.
It is in qualitative agreement with the phase diagram
of 3d simplicial gravity coupled to Ising spins~\cite{abjk}.
The shaded area reflects the uncertainty in the location of the
transition line separating the two phases of the geometrical system.
This uncertainty is due to a discrepancy between the results for
$\k_2^c$ arising when one uses different indicators for the change in
geometry. One possible indicator is the Hausdorff dimension, $d_H$,
another one the correlator $\la R^2 \ra - \la R \ra ^2$.
 The left boundary of the
shaded area results from determining $\k_2^c$ as the value of $\k_2$ at
the peak of $\la R^2 \ra - \la R \ra^2$.
The right boundary appears when $\k_2^c$ is
defined as the value of $\k_2$ for which there is a sudden change in
Hausdorff dimension. While the left hand boundary is relatively easy to
determine (Cf.\  figure~7) the right boundary is difficult to locate
precisely due to large fluctuations in geometry and should only be taken
as a rough estimate.
The fact that the two boundaries
do not coincide for the size of systems used here
should be taken as a clear sign of finite size effects. A related
phenomenon is seen in the numerical studies of 2d gravity coupled to
Ising spins, where the peak in the specific heat does not coincide
with the peak in the susceptibility due to finite size effects which
seem to disappear only very slowly when the size of the system is
increased.
The lines of phase transition (treating the shaded area as a ``line'',
which we expect it will be in the infinite volume limit)
divide the coupling constant plane into three regions: The one to the right is
characterized by no magnetization and elongated geometry, the
lower left region is characterized by no magnetization and
highly connected geometry, while the  upper left
corresponds to a magnetized phase and highly connected
geometry. It is difficult to determine the exact position of
the bifurcation point since we  have here both a fluctuating
geometry and large spin fluctuations. It is easy to understand that the
transition line separating different geometries  will
approach the value of $\k_2^c$ for pure gravity  when
$\b\to \infty$ and $\b \to 0$. In these limits the spin fluctuations
decouple from gravity and the locations of the transition must agree
with the one of pure 4d simplicial gravity.

In figure~5 we have shown the behaviour of the average curvature of our
manifolds when we fix $\k_2$ inside the highly connected phase fix and move
vertically in the coupling constant plane varying $\beta$. The value of
$\k_2$ is the same as in figure~1 and figure~2. The position of the peak
in the average curvature exactly coincides with the value of $\beta_c$
determined from the magnetization curve and the plot of Binders
cumulant. This observation allows an easy and not so time consuming
determination of $\beta_c(\k_2)$. The the transition line
$\beta=\beta_c(\k_2)$ was determined using this idea. We note that  this
line shows little dependence on $\k_2$. The dependence
of $\k_2^c(\beta)$ is more pronounced. The value of $\k_2^c$ is smaller
for the coupled system than for pure gravity. The shift in $\k_2^c$ is
largest when $\beta=\beta_c$ showing that the coupling between geometry
and spins is indeed largest when the spin system is critical.
This is in agreement with the intuition
we have from the exactly solvable 2d Ising-gravity system.
The transition line $\k_2=\k_2^c(\beta)$ shows
that effectively the spin system
pushes geometry towards larger $\k_2$ values. The effect is strongest
when $\b$ is close to $\b_c(\k_2)$. On the other hand we know that for large
$\k_2$ values the geometry is such that the system cannot be critical.
This apparent contradiction seems to be generic
for the interaction between gravity and matter of the kind considered
here. This is highlighted
in a recent paper on multiple spin systems coupled to 2d gravity
\cite{adjt}. In 2d the  {\it back-reaction} of the spin system on gravity is
also largest close to criticality, but is such that it counteracts its own
criticality by trying to deform the geometry into generic shapes
where it cannot be critical (polymer-like geometries). It seems that
we are observing a similar phenomenon here in 4d.

\vspace{12pt}

\noindent
It is of course an interesting question whether the coupling
between the spins and gravity changes the critical exponent of either of
the systems
as is the case in 2 dimensions. However, since
the critical exponents of the pure 4d gravity system are yet not known
and since it has proven quite difficult to extract by numerical
methods the critical exponents of the Ising spins coupled to
2d gravity, we have chosen here the more modest approach to look at
the influence of the spin system on bulk geometric quantities like
the average curvature. As explained in the introduction this
has special interest in relation to the scaling of gravity observables at the
transition between geometries. We will return to this aspect after we have
discussed briefly 4d gravity coupled to Gaussian fields.

\subsection{Gaussian fields coupled to gravity}

In the case of Gaussian fields we have, as explained above, no coupling
constant to adjust. The fields will automatically be critical in the
infinite volume limit. We have considered up to four Gaussian fields
coupled simultaneously to gravity and for these systems we can
make a statement similar to the one made for the Ising model:
The two phases of geometry seem to survive the coupling to
Gaussian matter. In fig. 6 we have shown the expectation value
$\la \phi^2 \ra$ of a single component of the Gaussian field as a function of
$\k_2$. We see a change in $\la \phi^2 \ra$ linked to the change in geometry.
The value of $\la \phi^2 \ra$ increases when
we enter into the elongated phase. In fact $\la \phi^2 \ra$ also has
quite large fluctuations in this phase.

\subsection{Behaviour of gravity observables coupled to matter}

In the computer simulations we can clearly see the back-reaction of matter
on the geometry for a given choice of coupling constants. It is
less obvious, however, that this back reaction of matter leads to
anything but trivial changes. Both for the coupling of Ising spins and Gaussian
fields we still have two phases of the geometry: the highly connected
one and the very elongated one. As mentioned in the introduction
one could hope that the inclusion of matter would improve the scaling
of the curvature at the transition. We have investigated this in the
following way: As remarked above there are several indicators of the
change in geometry. They result in slightly different values
of $\k_2^c$. We have chosen here to use the peak of
$\la R^2 \ra -\la R \ra ^2$
as an indicator of the transition, mainly because it is easier to
identify than the change in Hausdorff dimension.  The value of $\k_2^c$
depends on the matter content as can be seen from figure~7.
In fig.~8 we have plotted the average
curvature as a function of the distance $\Delta \k_2$ from
$\k_2^c$. It is seen that there is no improvement in the scaling
behaviour of $\la R \ra (\k_2^c)$ as a function of the matter content, when
we compare with the situation in pure gravity.
In fact the curves look remarkably insensitive to the inclusion of matter
and one could at this point wonder whether the back-reaction of
matter has any effect on the geometry except to  introduce an effective
$\k_2$ which differs from the bare parameter.
This is of course enough to explain the peak in the average curvature
observed in figure~5 and it also provides us with an explanation why the
peak is more narrow for a 9K system than for a 4K system. This is due to
the fact that the change in average curvature across the phase
transition is more sudden for the larger system.
In fig. 7 we have shown  $\la R^2 \ra -\la R \ra^2$ for various
matter fields coupled to gravity. We see that the peak grows
with the number of Gaussian fields, indicating at least somewhat increased
back-reaction with the number of fields. Furthermore we
note that the larger the number of Gaussian
fields is, the more $\k_2^c$ is shifted towards smaller values.
Hence systems with a large number of Gaussian fields favour elongated
geometries. The same phenomenon is known from two dimensions where
analytic considerations show that the path integral is dominated by
elongated geometries when $n_g$ is large.
However, there is no indication that the presence of matter fields
changes the nature of the phase transition of the geometrical system.

Let us comment here on a somewhat surprising feature of 4D simplicial
gravity. As mentioned earlier the method of grand canonical simulation
requires a finetuning of $\k_4$ to its critical value, $\k_4^c$.
It appears that $\k_4^c$ depends on $\k_2$ in a universal way. In
figure~9 we have shown $\k_4^c(\k_2)$ for pure gravity, gravity coupled
to Ising spins at $\beta=\beta_c$ and gravity coupled to 1 and 4
Gaussian fields respectively. In reference~\cite{ajk} 4D simplicial
gravity was simulated using the following action
\beq{sr2}
S=\k_4 N_4 -\k_2 N_2 + \frac{h}{c_4^2}
\sum_{n_2} o(n_2)\left(\frac{c_4-o(n_2)}{o(n_2)}\right)^2
\eeq
This corresponds to adding to the Einstein Hilbert
action a typical higher derivative
term (Cf.\ equation~\rf{*r2}). We have shown also $\k_4^c(\k_2)$ for this
model when $h=10$ and $h=20$. For all the systems
studied $\k_4^c(\k_2)$ is a linear function with a slope of
approximately $2.5$.

\section{Discussion \label{discuss}}

It is clear that the numerical exploration of simplicial quantum gravity
is still in its infancy. Finite size effects are not under control and it
would be most desirable to be able to simulate larger systems. In
principle it is possible and it {\it will} be possible in
the future. But even on  the small lattices used here one might
reveal interesting aspects of the interaction between gravity and
matter. Until now we
have only considered the simplest matter systems, spins and Gaussian fields,
but nothing prevents us from considering the coupling to for instance
non-abelian gauge fields. It is also in principle possible to
to define  non-local observables like spin-spin correlation functions
as functions of geodesic distance (see i.e. \cite{abjk} for a discussion
in the case of 3d gravity) and explore their quantum
averages.
In this paper we have not tried to extract
any critical exponents of such observables since the experience from
$3d$ is that it is not easy, and we decided in this first
investigation to concentrate on bulk quantities.

The main result of the simulations is that coupling of matter to
discretized gravity  seems {\it not} to influence the geometry in
a profound way. Of course it is possible that critical indices
change (as is the case in 2d gravity). Our measurements are still too
poor to measure such subleading effects. As mentioned above an interesting
effect would be an improved scaling of the average curvature in the region
where there is a transition in geometry. We have not seen any such effect.
The tentative conclusion from these first numerical experiments is that
matter fields (at least of the kind we have considered here) will not
add very much to our attempts to understand the basic structure of
four-dimensional quantum gravity.

\vspace{24pt}

\addtolength{\baselineskip}{-0.20\baselineskip}

\addtolength{\baselineskip}{0.20\baselineskip}

\newpage

\noindent {\large \bf Figure Captions}

\vspace{12pt}

\begin{itemize}

\item[Fig.\ 1] The absolute value of the magnetization, as defined
by \rf{*6}, as a function of $\b$ for $\k_2=0.9$, i.e. in the
phase with a highly connected geometry. The circles correspond to a
volume $N_4=4000$, the triangles to $N_4=9000$.

\item[Fig.\ 2] Binder's cumulant \rf{*6a} for $\k_2=0.9$ and three volumes:
$N_4=4000$ $(\bigtriangledown)$, $N_4=6000$ $(\Box)$ and $N_4=9000$
$(\bigcirc)$. The shape corresponds to a
transition of second or higher order and the  point of intersection
to $\b_c (N_4=\infty)$.

\item[Fig.\ 3] The absolute value of the magnetization, as defined
by \rf{*6}, as a function of $\b$ for $\k_2=1.3$, i.e. in the
phase with elongated geometry. The circles correspond to a
volume $N_4=4000$, the triangles to $N_4=9000$.

\item[Fig.\ 4] The phase diagram in the $(\k_2,\b)$ plane as it
appears when $N_4=9000$. As discussed in the text there are
reasons to believe that part of the diagram is distorted by
finite size effects and that the in the infinite volume the shaded
region will be replaced by the dashed line.

\item[Fig.\ 5] The effect on the curvature $\la R \ra-\la R \ra_0$
(where $\la R \ra$ is defined by \rf{*ac})
when we are in the phase with a large Hausdorff dimension and change $\b$.
The value of $\k_2 = 0.9$ and the circles correspond to $N_4=4000$
while the triangles correspond to $N_4=9000$. $\la R \ra_0$ denotes
the average curvature in the case of pure gravity (it differs slightly
for $N_4 =4000$ and $N_4=9000$ due to finite size effects).

\item[Fig.\ 6] The change in $\la \phi^2 \ra$
 (a single component field) as a function
of $\k_2$ for $N_4=4000$.

\item[Fig.\ 7] $\la R^2 \ra - \la R\ra^2$ for a different matter
fields as a function of $\Delta \k_2$.
Pure gravity $(\bigtriangledown)$, gravity + Ising at
$\beta_c$, $(+)$, gravity + 1 Gaussian field $(\bigcirc)$ and gravity +
4 Gaussian fields $(\Box)$. (The observables $\la R \ra$ and
$\la R^2 \ra $ are defined in~\rf{*ac} and~\rf{*r2} respectively.)

\item[Fig.\ 8] $\la R \ra$ as a function of $\Delta \k_2$ for different
matter content. Pure gravity $(\bigtriangledown)$, gravity + Ising at
$\beta_c$, $(+)$, gravity + 1 Gaussian field $(\bigcirc)$ and gravity +
4 Gaussian fields $(\Box)$.

\item[Fig.\ 9] $\k_4^c$ as a function of $\k_2$ for different systems.
Pure gravity $(\times)$, gravity + Ising at $\beta=\beta_c$
$(\bigcirc)$, gravity + 1 Gaussian field $(\Box)$, gravity + 4 Gaussian
fields $(\bigtriangleup)$, gravity with higher derivative term for
$h=10$ $(\bullet)$ and gravity with higher derivative term for $h=20$
$(+)$.
\end{itemize}
\end{document}